\documentclass{aa}  

\usepackage{graphicx}
\usepackage{txfonts}
\usepackage{lipsum}
\usepackage{subcaption}         
                                
\usepackage{lscape}            
                                
\usepackage{placeins}

\begin{document}

   \title{Globular cluster candidates uncovered by VVVX and MUSE}

	\author{P. Rodríguez Cuevas\inst{1}\fnmsep\thanks{p.rodrguezcuevas@uandresbello.edu} \and 
    B. Dias\inst{1} \and 
    D. Minniti\inst{1,2} \and
    E. Garro\inst{3} }

	\institute{Instituto de Astrof\'isica, Departamento de F\'isica y Astronom\'ia, Facultad de Ciencias Exactas, Universidad Andres Bello, Fernandez Concha, 700, Las Condes, Santiago, Chile. \and
    Vatican Observatory, Vatican City State V-00120, Italy
    \and
    ESO - European Southern Observatory, Alonso de Cordova 3107, Vitacura, Santiago, Chile.}

   \date{Received: ; Accepted: }

  \abstract
   {The Milky Way has about 220 cataloged globular clusters. This number is far smaller than what is found in a similar galaxy such as Andromeda, which has about 600 globular clusters. Two possible reasons for this large difference might be observational bias and cluster dissolution. The infrared photometric survey VISTA Variables in the Via Lactea Extended has discovered many globular cluster candidates toward the Galactic bulge that need confirmation. 
   }
   {We confirm the globular cluster nature of four candidates from the Galactic bulge based on radial velocities from MUSE data.} 
   {Following our previous experience and method, we searched for a peak in the radial velocity distribution by deconvolving the radial velocity distribution into its bulge, disk, and potential cluster components.}
   {For Minni\,23, we confirm the previously found parameters: this is an old and metal-rich bulge globular cluster. Additionally, we found a radial velocity peak associated with the candidate globular cluster Minni\,07, with $\mu_{RV,cl}=80^{+26}_{-22}$ ${\rm km\ s}^{-1}$, $\sigma_{RV,cl}=48^{+15}_{-15}$ ${\rm km\ s}^{-1}$. We also derived $A_V=1.5$ and [Fe/H]$\approx-0.27$ for Minni\,07. On the other hand, we were unable to reliably confirm the globular cluster nature for the candidates Minni\,10 and Minni\,19 with the currently available data.}
   {The devised strategy has proven to be efficient for the proposed goals, and it is worth repeating it for the larger sample of candidate globular clusters.}


   \maketitle
   \nolinenumbers

	\section{Introduction}
    Globular clusters (GCs) contain stellar populations that can be as old as the Milky Way (MW). This makes them interesting objects that can help us to study the chemical and structural evolution of galaxies. For different galaxies, we can find different numbers of GCs. This number has even shown a correlation with the luminosity or baryonic mass of the galaxy \citep{Harris+2013}. However, in the case of GCs found in the MW, their number has been much smaller than what is expected. It reached 157 cataloged GCs as of 2010 in the latest version of the \citet{harris1996} catalog, which is smaller than that of Andromeda, which reaches almost 600 GCs \citep{Barmby&Huchra+2001,Huxor+2014}. This number is much larger, even though the two galaxies have similar masses. These differences between the MW and Andromeda might arise from our inefficiency in detecting GCs in the MW. These missing GCs might be low-mass clusters that were disrupted in the past, making them hard to detect, since their luminosity would be low (and would therefore not change the luminosity or mass of the MW significantly), and they might lie in regions of the MW that are difficult to observe, such as regions close to the Galactic plane.	Recent efforts to discover new GCs have been summarized in two recent compilations by \citet{bica+2024,garro+2024}, which contain more than 200 GCs in the MW. This is still well below the $\sim$600 GCs from Andromeda.
    
	The MW bulge is the most complex region to observe, especially close to the Galactic plane, because of the very high crowding and high-differential reddening. These cause difficulties in studies with optical observations and make the bulge one of the least studied regions of our Galaxy. However, near-infrared (NIR) photometric surveys such as the Two Micron All Sky Survey (2MASS; \citealp{Skrutskie+2006}) or the Vista Variables in the Via Lactea (VVV and its extension VVVX; \citealp{Minniti+2010}) have allowed us to further investigate in these high-crowding regions, in which the reddening has a lesser effect in the NIR than in the optical bands. The possibility of finding low-mass GCs is plausible in the bulge because stars in the Galactic bulge are old, and the bulge is a region in which many dynamical processes have occurred, such as dynamical friction, bulge shocking, evaporation, and tidal disruption that could have disrupted several GCs and reduced them to low-mass GCs \citep{Minniti+2017a}.
    
    Surveys such as the VVV now enable us to search for these missing clusters of the MW in the Galactic bulge. Since GCs are old stellar populations, some of the ways to search for them is finding overdensities of tracers of old populations in small regions, such as RR Lyrae, red clumps, red giant stars, or type II Cepheids. With VVV data and the search for these overdensities, a total of 350 GC candidates have been found in the Galactic bulge region and are named Minni clusters \citep{Minniti+2017a,Minniti+2017b,Minniti+2017c,Minniti+2019,Minniti+2020,Minniti+2021}. Currently, several analyses have been made to determine whether these are actual GCs through photometric data, proper motions, and distances. The conclusions were mixed. \cite{Piatti+2018} used 2MASS data and reported that Minni 01 to 22 were actually bulge GCs. However, \cite{Gran+2019} combined data from VVV and Gaia DR2 to analyze Minni\,01 to Minni\,84 and concluded that none of them is a GC. These two conclusions cannot be valid simultaneously, and deeper studies are therefore required to determine the nature and properties of the Minni clusters.
    
    The radial velocity (RV) is a powerful information that can be used to verify the existence of these GC candidates. The RV distribution, in combination with spectroscopic metallicities and multiband photometry, was analyzed before and was able to confirm one of the Minni clusters, more specifically, FSR\,1776 or Minni\,23 \citep{Dias+2022}. This cluster was already considered a promising candidate by \cite{Palma+2019}, who analyzed its CMD. The objective of this work is to use MUSE-based RVs from an initial sample of four Minni cluster regions in order to search for a RV peak other than that from the bulge and disk stars. When a RV peak is detected, we characterize the cluster candidate using MUSE-based spectroscopic metallicities and multiband photometry using a method adapted from \cite{Dias+2022}. The clusters we studied were Minni\,07, Minni\,10, Minni\,19, and Minni\,23.

This paper is organized as follows. In section 2 we describe the target selection and how we obtained the data. Section 3 describes the radial velocity analysis, models, and fits. Section 4 is dedicated to the spectroscopic and photometric analysis of Minni 07. Finally, our conclusions are summarized in Section 5.

	\section{Target selection and data}

    \subsection{Spectroscopy}

    From the hundreds of new bulge GC candidates selected using VVVX survey data, we selected 30 candidates with lower extinction that were to be observed in visible wavelengths with the Multi Unit Spectroscopic Explorer (MUSE, \citealp{Bacon+2010}) on the 8m UT4 from the Very Large Telescope (VLT) as a spectroscopic follow-up. The data were acquired during 2019 and 2020 under projects 0101.D-0363(A) and 0103.D-0546(A) (PIs: Minniti and Dias, respectively). The first MUSE datacube on Minni23 was analyzed and published by \citet{Dias+2022}. For this paper, four Minni clusters were chosen (see Table \ref{table:1}). They were selected based on the image quality and by observing their RV distribution, and we searched for those whose peak might be associated with a cluster based on an initial visual inspection alone. These clusters will be used to test a method that was improved since \citet{Dias+2022} before the whole sample is analyzed in a subsequent paper.

	\begin{table}[!htb]
		\caption{Globular Clusters log of observations} 
		\label{table:1}
        \footnotesize
		\centering
		\begin{tabular}{c@{\hskip 0.2cm}c@{\hskip 0.2cm}c@{\hskip 0.2cm}c@{\hskip 0.2cm}c@{\hskip 0.2cm}c} 
			\hline\hline 
			ID & l° & b° & Seeing & FWHM & Obs. date  \\ 
			\hline 
			Minni 07 & -2.58 & -5.46 & 0.87" & 0.52" & 15/May/2019  \\ 
			Minni 10 & -7.46 & -3.86 & 1.02" & 0.62" & 08/Jun/2019  \\    
			Minni 19 & -4.85 & -1.71 & 0.69" & 0.48" & 12/May/2019  \\     
			Minni 23 & -5.28 & -5.25 & 1.11" & 0.74" & 08/Jun/2019 \\   
		\hline 
		\end{tabular}
	\end{table}

    Observations were performed in service mode with MUSE in AO mode with a wide field of 1'x 1'. The data were reduced with the ESO standard pipeline during phase III. The natural seeing varied from $\sim0.7\arcsec$ to $\sim1.1\arcsec$, and the adaptive optics improved the PSF to an FWHM from $\sim0.5\arcsec$ to $\sim0.7\arcsec$, i.e., $\sim65$\% of the real seeing on average.

    The spectra of the stars were extracted with PampelMUSE \citep{Kamann+2013,Kamann+2018} using a point source catalog generated from the PSF photometry of the MUSE convolved image as input. The RVs were estimated through a cross-correlation in the optical region using templates from the MILES library \citep{Sanchez+2006} with a spectral type similar to the observed stars found using the ETOILE code \citep{Katz+2011,Dias+2015}. The RVs were also estimated using the near-infrared region of the spectra and reached very similar values. The RV errors were estimated using the relations derived by \cite{Valenti+2018} by performing Monte Carlo simulations as a function of S/N and [Fe/H]. The S/N was obtained from the PampelMUSE simplified routine for this task as first guess, and the [Fe/H] came from the full spectrum fitting of atmospheric parameters performed with ETOILE. For more details on this data analysis, we refer to \citep{Dias+2015, Dias+2022}.

    \subsection{Photometry}

For the subsequent analysis, we used the source catalog from MUSE and matched it with three photometric catalogs: 
    (i) the VVV survey DR5 PSF photometry \citep{alonso-garcia+2018} for filters J and Ks, (ii) the DECaPS DR2 \citep{schlafly+2018} for filters g, r, i, z, and Y, and (iii) Gaia DR3 \citep{gaia2016,gaiaDR3} for filters G, BP, and RP.

    \section{Radial velocity analysis}
    \subsection{Procedure}
	The way to search for the cluster through RVs is by observing their RV distribution. Since the observed stars in our line of sight will not only have cluster stars, but also foreground stars from the MW disk and background stars from the bulge, we expect to find the RV distribution for these two components in our observations. When we consider that GC are a group of gravitationally bound stars that move together,  we expect the kinematic parameters such as distance, proper motions, and RVs to be very similar among the cluster stars. Therefore, a third component in the RV distribution that is different from the disk and bulge would be observed as a distinct peak in the RV histogram. When this component is found, it is strong evidence toward confirming the cluster nature of the Minni cluster candidate. However, it is not necessarily straightforward to find this component because the kinematic parameters can overlap the disk and bulge component \citep{Garro+2021}.

\begin{figure}
		\centering
		\includegraphics[width=8cm]{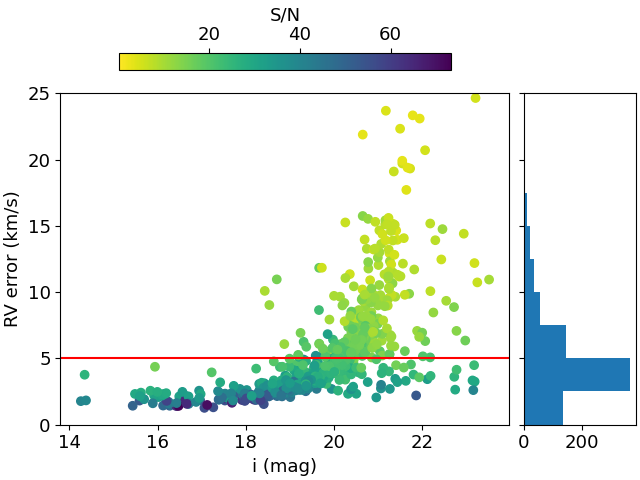}
		\caption{RV error of Minni\,07 as a function of i magnitude. The color bar shows the S/N for each observation. The red line marks the error cut of 5 ${\rm km\ s}^{-1}$ for the target selection. The right histogram shows the distribution of RV errors.}
        \label{fig:RVerror}
	\end{figure}

\begin{figure}
		\centering
		\includegraphics[width=8cm]{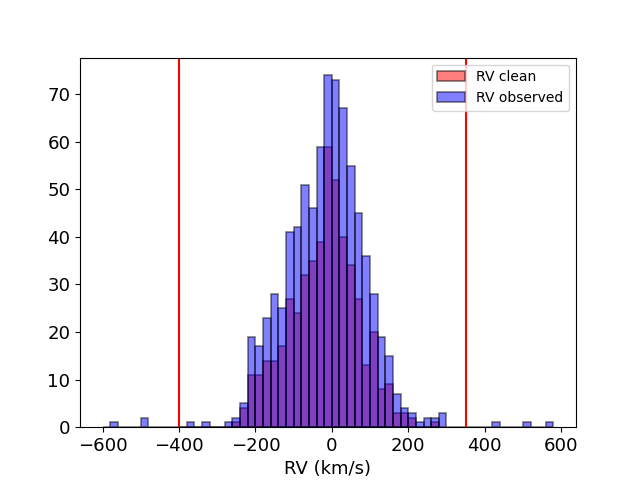}
		\caption{Observed RV distribution of Minni\,07 before (blue) and after (red) we removed stars with $RV_{error} > 5$ ${\rm km\ s}^{-1}$. Stars outside the range defined by the red vertical lines were considered outliers and removed.
        }
        \label{fig:RVcleaned}
	\end{figure}

The first step in our RV analysis was to select high-quality spectra. We used Figure \ref{fig:RVerror} with the RV error as a function of stellar magnitude and color-coded by the S/N and decided to make a cut in RV error to be smaller than the typical GC velocity dispersion. The dispersion velocity of GCs grows with their mass and reaches values ranging from 1 to 20 ${\rm km\ s}^{-1}$ \citep{Norris+2014}, such as $\omega$ Centauri with $\sigma=22$ ${\rm km\ s}^{-1}$ for the most extreme case \citep{Noyola+2010}. The typical dispersion is about $\ 5$ ${\rm km\ s}^{-1}$, and the Minni clusters are expected to have a lower mass, which means relatively smaller velocity dispersions. Our simple cut was $RV_{error} <5$ ${\rm km\ s}^{-1}$. 
The second step was to plot the cleaned RV distribution and remove potential outliers with a visual criterion (see Figure \ref{fig:RVcleaned} for the case of Minni 07). The statistics of the selected stars can be found in Table \ref{table:specstat}.

\begin{table}[!htb]
		\caption{Statistics of the extracted spectra per MUSE datacube.} 
		\label{table:specstat} 
        \small
		\centering 
		\begin{tabular}{c c c c}
			\hline\hline 
			ID & $ N_{all\ spec}$ & $ N_{good\ spec}$ & $ Mag_{lim}^{(a)}$ \\ 
			\hline 
			Minni\,07 & 802 & 500 (62\%) & 20.4 \\
			Minni\,10 & 689 & 224 (32\%) & 20.4 \\ 
			Minni\,19 & 865 & 461 (53\%) & 20.6 \\ 
			Minni\,23 & 450 & 297 (66\%) & 19.8 \\ 
		\hline 
		\end{tabular}
    \tablefoot{
    $ ^{(a)}$ This magnitude is a rough estimate of the relative comparison of the depth of each MUSE datacube. It was calculated as the median magnitude of all stars with $4.5 < \sigma_{ RV} ({\rm km\ s^{-1}}) < 5.5$.}
	\end{table}

\begin{figure}
		\centering
		\includegraphics[width=8cm]{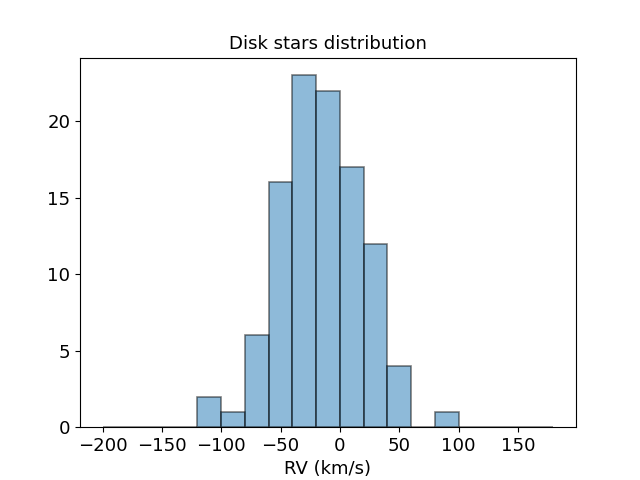}
		\includegraphics[width=8cm]{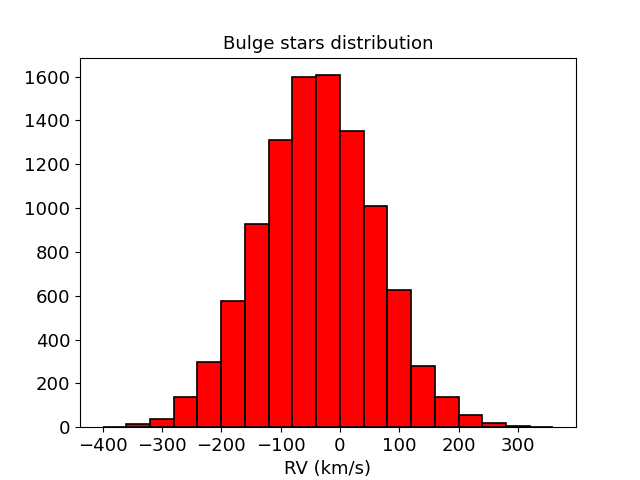}
		\caption{RV distribution for the disk and bulge obtained with Gaia DR3 and GIBS for Minni\,07.}
        \label{fig:RVdiscbul}
	\end{figure}

	In order to determine whether a component was associated with the cluster, we simulated the RVs for disk and bulge stars to compare the model of field stars with the cleaned RV distribution from the MUSE observations. The bulge was modeled using the GIBS survey interpolator \citep{Zoccali+2014}, obtained using red clump stars in different bulge fields. The interpolation uses Galactic coordinates (l,b) to give the values of the average RV velocity $\mu_{RV}$ and the dispersion $\sigma_{RV}$ for the bulge in galactocentric RVs, which were later transformed into heliocentric RVs (Figure \ref{fig:RVdiscbul}) so that they could be compared to our observed data. For disk stars, we used Gaia data release 3 \citep{Gaia+2016,Gaia+2023}, from which we extracted the targets by selecting Minni cluster positions with a radius of 10 arcminutes (this radius is about ten times the size of each of our cluster fields of view). Because the disk has a younger population, we used a color-magnitude diagram from Gaia photometry to select young bright main-sequence stars that represent the disk population well and estimated the values of $\mu_{RV}$ and $\sigma_{RV}$ for the RV of the disk component. Additionally, the bulge parameters for RVs were reestimated again using Gaia DR3, but by searching for the older population (RGB stars). The results were similar to those calculated with the GIBS survey interpolator for the four clusters.
    
\begin{figure}
		\centering
		\includegraphics[width=8cm]{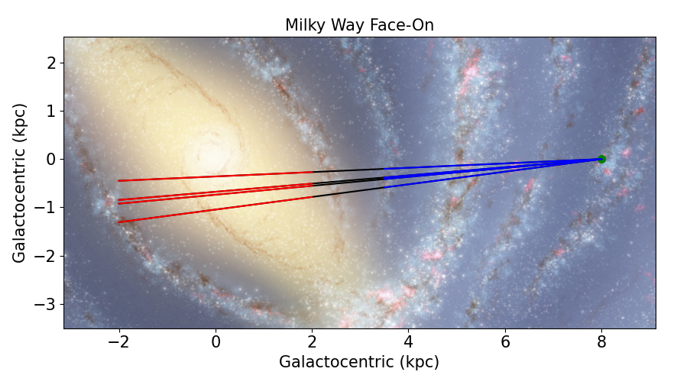}
		\caption{Selection of disk and bulge stars for each direction of the observed regions. The green point shows the Sun, and stars at a distance $<4.5$ kpc were considered to be part of the disk (blue region). Stars at a distance $>6$ kpc from the Sun were considered to be part of the bulge. Stars in the transition region (shown in black) were ignored.}
        \label{fig:MW separation}
	\end{figure}

    The proportion between disk and bulge stars in our line of sight is required, so that the model can represent the RV distribution of field stars as reliably as possible. To achieve this, we used Besançon models for each of the cluster directions, taking all stars with a distance from the Sun d<4.5 kpc as disk stars and d>6 kpc as bulge stars. Stars between 4.5 kpc and 6 kpc were ambiguous and we did not consider them to calculate the stellar proportion (see the illustration of our criterion in Figure \ref{fig:MW separation} ). After we estimated the relative population of disk and bulge stars, we created a model that only used the distributions of the disk and bulge, assuming a Gaussian distribution, and with the same numbers of stars as were found in the Minni clusters regions, following the stellar proportion estimated in the Besançon model.

    The next step was to compare the RV distribution of the bulge+disk model with the observations. To do this, we used the Kolmogorov-Smirnov test (K-S test), whose p-value tells us whether these two distributions come from different samples. Since the comparison model only includes the RV distribution of disk and bulge stars, a third component would be included in the RV distribution separated from the disk and bulge if there were a cluster. The result would be that the field stars model and observation are statistically different. In order to detect a GC, we therefore first required that the K-S test returned a low p-value $<$0.05 to reject the null hypothesis, that is, to confirm that the bulge+disk stars model does not represent the observations well.
    
\begin{figure}
		\centering
		\includegraphics[width=8cm]{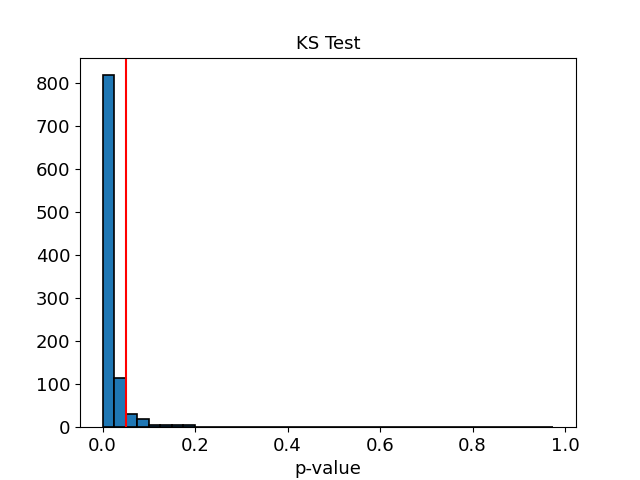}
		\caption{Bootstrapping results showing the p-value for a 1000 K-S test between the bulge+disk model vs. observation for Minni\,07. The red line shows the p-value$=0.05$.   
        }
        \label{fig:KStest}
	\end{figure}

	To alleviate the stochastic effects imprinted on the model sampling, we performed a bootstrap strategy, resampled the simulated RV distribution 1,000 times, and ran the K-S test for each simulation.
    Figure \ref{fig:KStest} shows the distribution of the 1,000 K-S test results for the case of Minni\,07, showing that in about $93\%$ of the cases a p-value$<$0.05. In conclusion, stochastic effects do not change the result, and we confirm that the RV distributions of the bulge and disk alone cannot represent the observed RV distribution. There must be an additional RV contribution from another component.

	In order to find a GC in an RV distribution, we needed to fit a third component associated with that GC. Based on the assumption that it follows a Gaussian distribution, the parameters of this third component will have an average RV $\mu_{RV,cl}$ and a velocity dispersion $\sigma_{RV,cl}$. However, the fraction of stars that are part of the cluster in our line of sight also needs to be taken into account, so there is a third parameter related to the number of cluster stars that also needs to be fit $frac_{cl}$. A natural method for this endeavor would be a Gaussian mixture model. However, \citet{Dias+2022} demonstrated through simulations that this method is not suitable for this particular analysis.
    
\begin{figure}
		\centering
		\includegraphics[width=8cm]{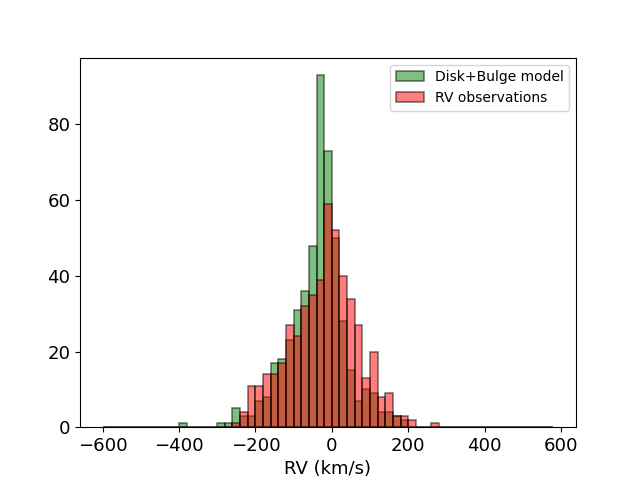}
		\caption{Simulated RV distribution for Minni\,07 for bulge+disk stars (green), and MUSE observations after removing outliers and high RV error (red, as in Fig.\ref{fig:RVcleaned}).}
        \label{fig:Prior Minni 23}
	\end{figure}

	The fitting method we used to find the cluster parameters was the Markov chain Monte Carlo (MCMC) method. MCMC determines $\mu_{RV,cl}$, $\sigma_{RV,cl}$, and $frac_{cl}$ by comparing the RV distribution model taken from the bulge, disk, and cluster stars with the observed RV distribution. All the kinematic parameters and stellar proportion of the disk and bulge components remained fixed, and only the parameters of the third component varied. We note that the relative size between the third component sample and the bulge+disk sample was adapted for each third component fraction selected by a given walker in the parameter space. Because we compared two histograms, we assumed that the model and observation bin heights followed a Poisson distribution, giving the likelihood calculated by \cite{Tremmel+2013}, and adapted by \cite{Bernardo+2024},    
	\begin{equation}
	\prod_{Bins}\frac{\Gamma(0.5 + n_{obs} + n_{model})}{\Gamma({1+n_{obs}}) \Gamma({n_{model}+0.5})},
	\end{equation}	
    
	\noindent in which $n_{obs}$ and $n_{model}$ are the bin heights, and the binning was the same in the two histograms. Finally, taking the logarithm, we obtained the likelihood for our MCMC,	
	\begin{multline}
	Likelihood=\sum_{Bins}\log(\Gamma(0.5 + n_{obs} + n_{model})) \\
	- \log(\Gamma({1+n_{obs}})) - \log(\Gamma({n_{model}+0.5})).
	\end{multline}
    The prior we selected for MCMC was a grid of points between $0<\sigma_{RV,cl}<40$ ${\rm km\ s}^{-1}$ and $0.03<frac_{cl}<0.3$. In the case of $\mu_{RV,cl}$, the ranges were selected by comparing the histogram of the bulge+disk stars model against the MUSE RV observations and searching for the region in which we expected to find the third component. For instance, Figure \ref{fig:Prior Minni 23} shows that for Minni 07, the third component might be around $100$ ${\rm km\ s}^{-1}$, so the ranges for the prior in Minni 07 were $40<\mu_{RV,cl}<200$ ${\rm km\ s}^{-1}$. 
    
    \subsection{Results}

    All the steps of the analysis process described in the previous section were applied to the four clusters Minni 19, Minni 23, Minni 10, and Minni 07. For Minni 19 (see the appendix), we found that the K-S test mainly returned a p-value>0.05. This means that we were unable to determine a third component for Minni 19 that is associated with the GC because the disk and bulge stars fit the observations well. However, this does not mean that the cluster does not exist. The RV distribution of the cluster might overlap very well with either the disk or the bulge.
    
	The cluster Minni 23 has been studied before in \citet{Dias+2022}. We found using the K-S test with the clean sample that the p-value$<$0.05 is close to $\sim$80\% of the bootstrap simulations. We therefore rejected the null hypothesis, and the majority of the samplings of the models did not represent the observations well. A third component might be found in RV observations. Applying MCMC, we searched for the parameters assuming that the third component is a Gaussian, as mentioned before (see the appendix), resulting in $\mu_{RV,cl}=-115^{+15}_{-16}$ ${\rm km\ s}^{-1}$, $\sigma_{RV,cl}=32^{+12}_{-11}$ ${\rm km\ s}^{-1}$, and $frac_{cl}=0.17^{+0.07}_{-0.06}$. The parameter values were the median taken from the MCMC, and the error bars are the 16\% and 84\% percentiles. The results are similar to those from \cite{Dias+2022}, $\mu_{RV,cl}\approx-110$ ${\rm km\ s}^{-1}$, $\sigma_{RV,cl}\approx32$ ${\rm km\ s}^{-1}$, and $frac_{cl}\approx0.12$.

\begin{figure}[!htb]
    \centering
    \includegraphics[width=\columnwidth]{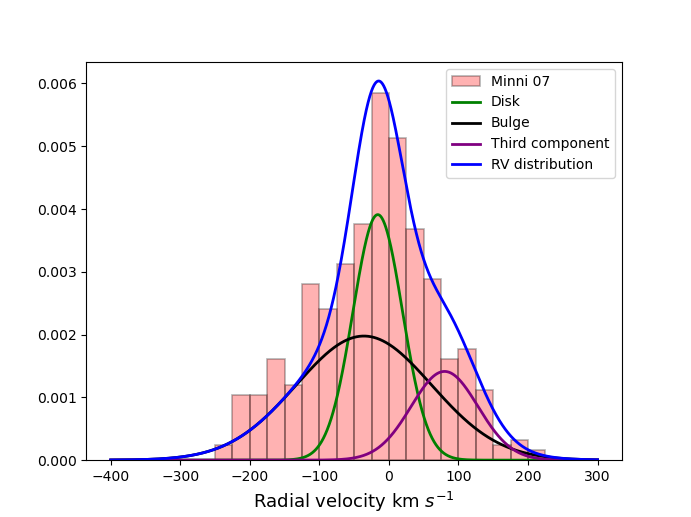}
    \caption{Radial velocity distribution for Minni\,07, with the results of the three-Gaussian fit.}
    \label{fig:RV3gauss}
\end{figure}

	For Minni\,07, $\sim$93\% of the K-S tests with the 1,000 bootstrap simulations gave a p-value$<$0.05 (Figure~\ref{fig:KStest}). Therefore, just as for Minni\,23, we searched for the third component using an MCMC, which resulted in $\mu_{RV,cl}=80^{+26}_{-22}$ ${\rm km\ s}^{-1}$, $\sigma_{RV,cl}=48^{+15}_{-15}$ ${\rm km\ s}^{-1}$ and $frac_{cl}=0.13^{+0.07}_{-0.05}$, as shown in Figs. \ref{fig:RV3gauss} and \ref{fig:mcmc Minni23}. 

    Finally, for Minni 10,  $\sim$90\% of the cases gave p-values$<$0.05 (see the appendix). Since the field stars model fit the observations only poorly, we attempted to find a third component using MCMC. Unfortunately, we were unable to find a set of parameters for the third component because the parameters did not converge to a result. We note that Minni 10 lost a great number of stars after the error cut, unlike the other three Minni clusters.

    We remark that the likelihood we used for MCMC compares the bin height of the models and observations. The bin width was chosen following the Freedman-Diaconis rule, leading to $\sim$ 20 ${\rm km\ s}^{-1}$. We tested bin widths of 15 and 30 ${\rm km\ s}^{-1}$ to check the sensitivity of the results on the bin width choice. The final results were within much less than one $\sigma$ with respect to the original results. Therefore, we kept 20 ${\rm km\ s}^{-1}$ as the bin width for the clusters in this paper.
    
    We also tested the effect of the RV errors on the MCMC results. The analysis was made based on stars with uncertainties lower than 5 ${\rm km\ s}^{-1}$, which is four times lower than the bin width. The histogram distribution therefore did not change significantly when we dispersed the measurement based on the individual RV error. In other words, the bin height, which is the basis of the likelihood described above, did not change drastically.
    
    The source of the large parameter uncertainties might be a low number statistics effect combined with possible systematic errors from the models. Future observations and models might help with this matter.

	\begin{figure}
		\centering
		\includegraphics[width=8cm]{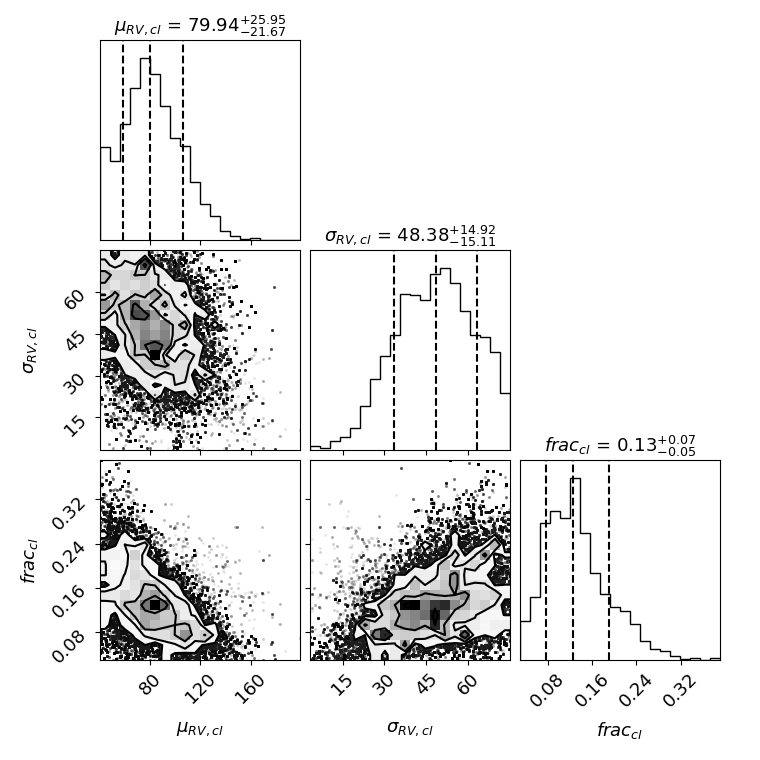}
		\caption{MCMC for parameters of the cluster component for Minni\,07.}
        \label{fig:mcmc Minni23}
	\end{figure}

    \section{Characteriziation of Minni\,07}

    After we confirmed the third component in the RV distribution for Minni\,07, we characterized this cluster candidate. We used the multiband photometry from VVV, DECAPS, and Gaia to analyze the color-magnitude diagrams (CMDs) with isochrones. Four parameters are required to define the isochrone: age, metallicity, distance, and extinction. We adopted the PARSEC isochrones \citep{bressan+12}, and they provided the extinction coefficients for all filters from these surveys following \cite{Cardelli+1989} and \cite{O'Donnell+1994}.

\begin{figure*}[!htb]
    \centering
    \includegraphics[width=\textwidth]{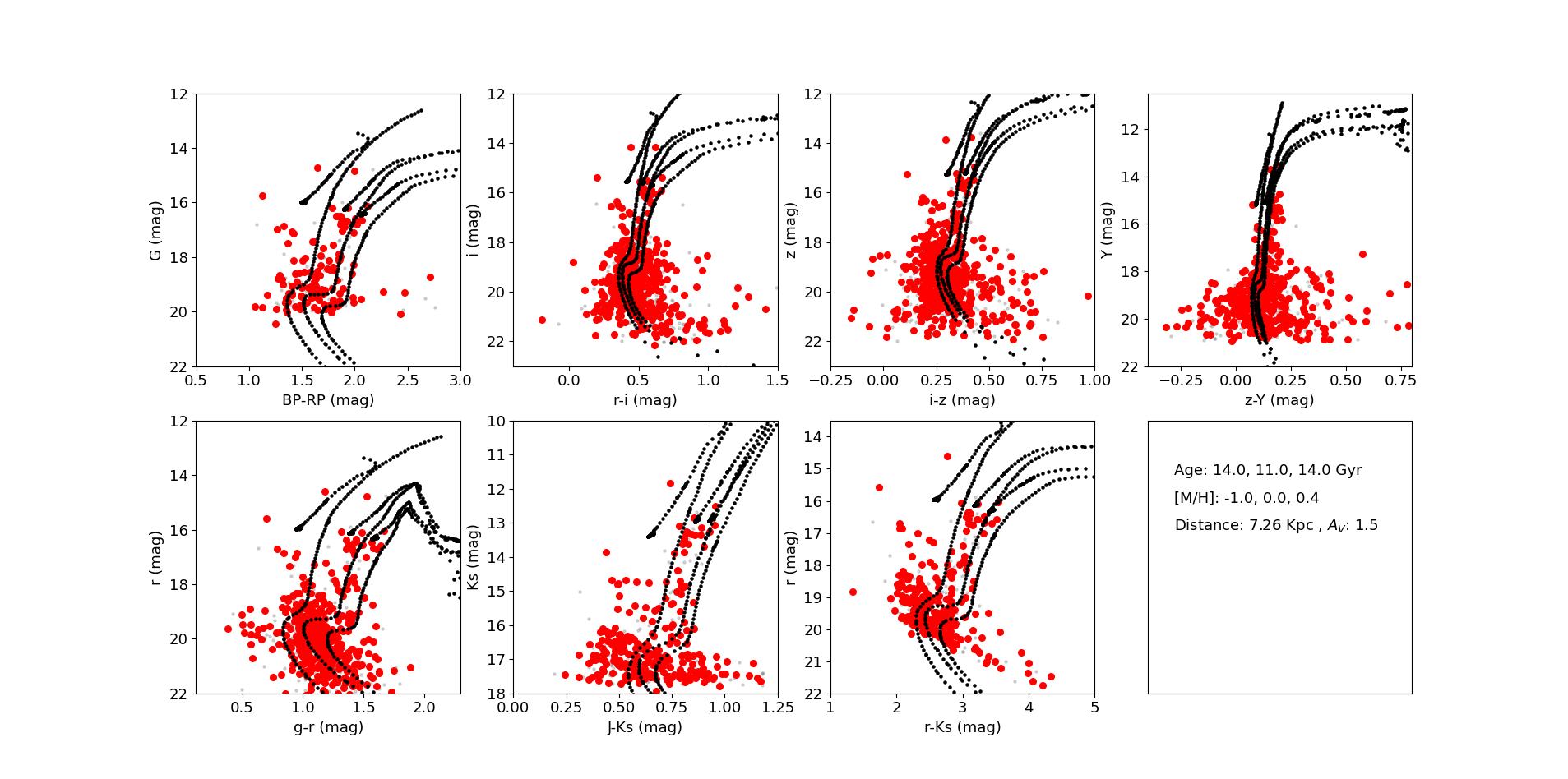}
    \caption{Multiband CMDs for bulge stars.
    The red points show bulge stars in the field of Minni 07 selected within $\mu_{RV,bulge}\pm1\sigma_{RV,bulge}$. We show PARSEC isochrones with parameters from \cite{clarkson+2008} and $A_V=1.5$. The three isochrones from left to right correspond to the pairs of age and metallicity listed in the figure from left to right. The distance and extinction were kept fixed.} 
    \label{fig:bulgecmd}
\end{figure*}

To estimate the value of $A_V$, we used the same isochrones that \cite{clarkson+2008} used to describe the bulge as a reference to fit our bulge selection. 
To select bulge stars, we used their RVs and only selected stars with RVs within $\mu_{RV,bulge}\pm1\sigma_{RV,bulge}$. Using these isochrones by visual inspection, we only varied $A_V$ and found that $A_V=1.5$ fits the bulge CMD well (Fig. \ref{fig:bulgecmd}). Our visual fit sought to wrap the stars around the red giant branch and the main-sequence turn-off with the three isochrones in all bands. This extinction translates into E(J-Ks)=0.23, following the same extinction law as we used to fit the multiband CMDs altogether. This value is consistent with E(J-Ks)=0.21 found by \cite{Minniti+2017c}. We kept this extinction fixed for the cluster candidate analysis. For the distance, we assumed 6.8\,kpc derived by \cite{Minniti+2017c} from IR CMD of the VVV, which is robust for a distance determination. In this way, we fixed two out of four parameters, which reduced the degeneracy in the following analysis.

\begin{figure*}
	\centering
	\includegraphics[width=0.45\textwidth]{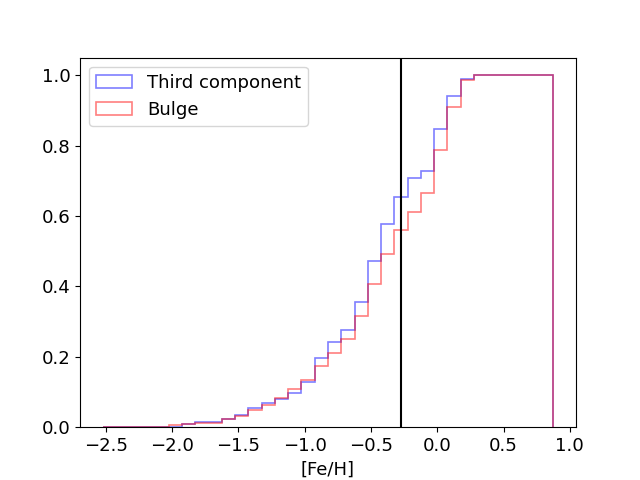}
    \includegraphics[width=0.45\textwidth]{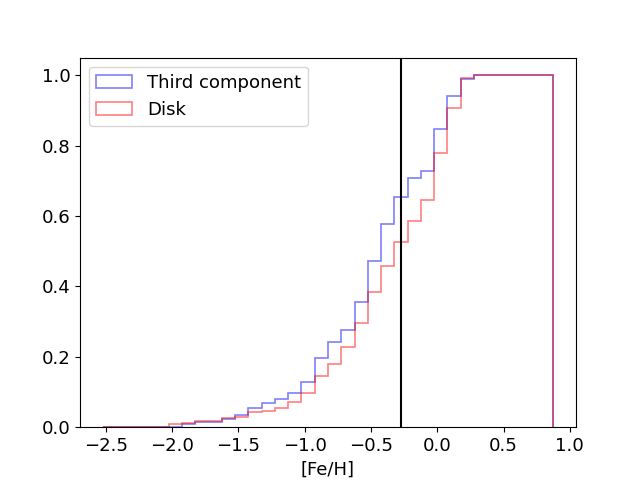}
	\caption{Cumulative distribution of [Fe/H] for the bulge and third component (left). Cumulative distribution of [Fe/H] for the disk and third component (right). They both show a major difference around [Fe/H]$\approx-0.27$. This value was taken for the third component isochrone.}
    \label{fig:minni07met}
	\end{figure*}

In order to estimate the metallicity of Minni\,07, we compared its metallicity distribution with the distributions from the bulge and the disk stars and searched for contrast. The bulge, disk, and Minni\,07 stars were selected within $\mu_{RV}\pm1\sigma$ from their respective RV peaks. For this particular field, the bulge and disk significantly overlap each other, which makes the cross-contamination very high. Their [Fe/H] distributions are therefore very similar with mixed populations.
We compared the cumulative [Fe/H] distribution of the cluster with that of the bulge (or disk) using the K-S test.
We found a statistically significant difference with p-values of 0.005 and 0.008 for the bulge and disk, respectively. The position of the maximum difference was at [Fe/H]=-0.2735 in both cases, and we assumed this as the Minni\,07 metallicity (see Fig. \ref{fig:minni07met})

\begin{figure*}[!htb]
	\centering
	\includegraphics[width=\textwidth]{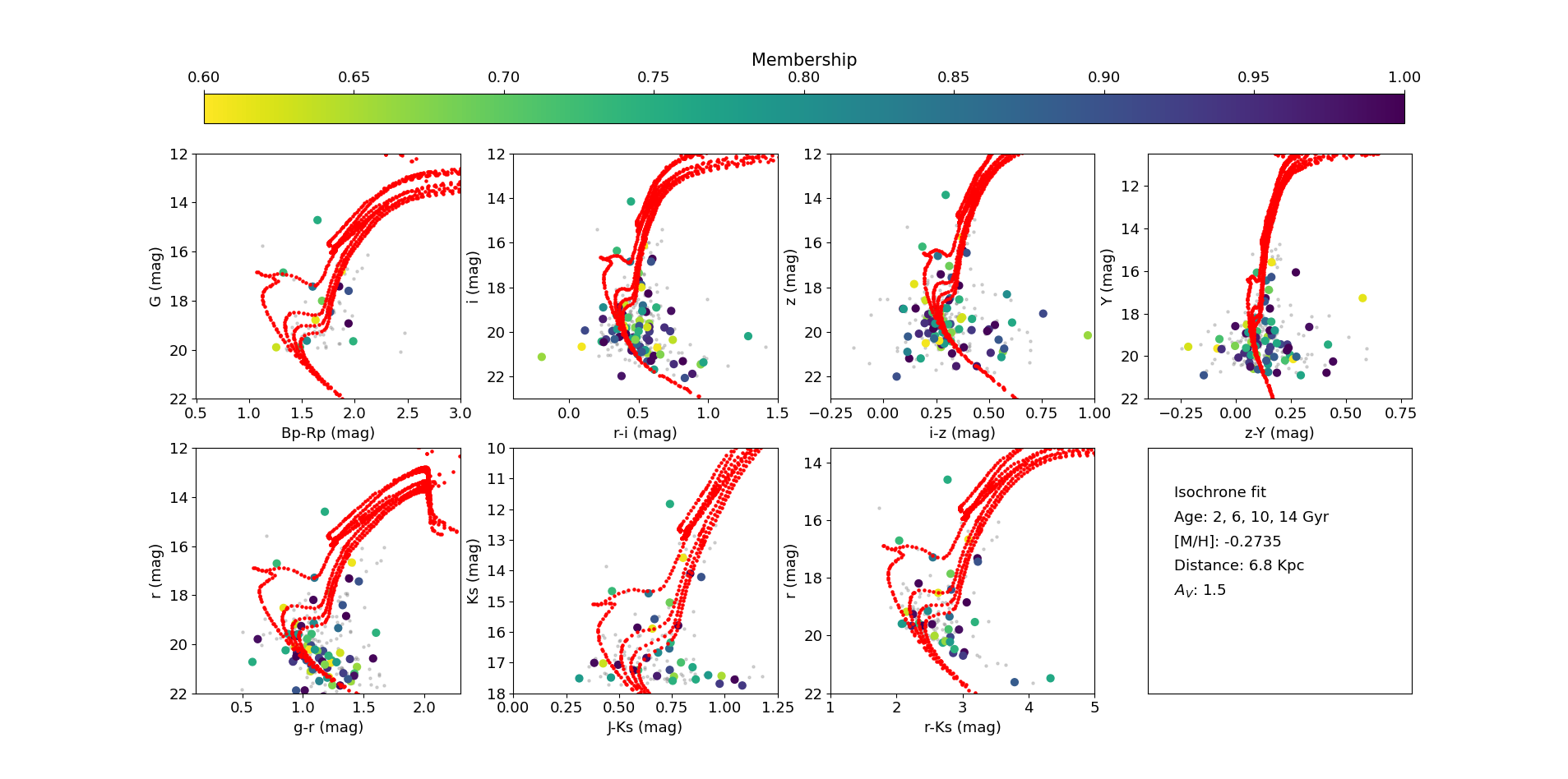}
	\caption{Multiband CMDs for Minni\,07 stars.
    The colored points are Minni\,07 third-component stars selected within $\mu_{RV,cl}\pm1\sigma_{RV,cl}$ and $-0.5<$[Fe/H]$<0.0$. The gray points are Minni 07 stars that were not selected. The color code defined as membership means the normalized height of the RV Gaussian fitted to Minni\,07.
    We show PARSEC isochrones with ages 2, 6, 10, and 14 Gyr from top to bottom, with fixed $A_V=1.5$, 6.8\,kpc and [Fe/H]=-0.27. }
    \label{fig:minni07isoc}
	\end{figure*}

    With this metallicity found together with the distance from \cite{Minniti+2017c} and estimated $A_V$, we fitted the isochrones in the CMD for Minni\,07 with different ages. The Minni\,07 stars selection was based on metelliticy around the value estimated above of [Fe/H]=-0.2735, within the range $-0.5<$[Fe/H]$<0.0$, in addition to the previous RV criterion of $\mu_{RV}\pm1\sigma$. 
    After trying different ages from 2 to 14 Gyr, we found that the third component likely is an old population (see Fig.\ref{fig:minni07isoc} with the best visual fit).
    
	\section{Summary and conclusions}

    Four GC candidates found in the MW bulge with VVV were selected and analyzed through their RVs obtained with MUSE. The clusters were searched for by comparing their respective RV distributions with a simulation that only considered disk and bulge stars contributions in the direction of that particular field.
	We recreated the results found by \citet{Dias+2022} for Minni\,23 and found a third component in the RV distribution with similar values using an MCMC method. For Minni\,07, we found the parameters for a third RV component with $\mu_{RV,cl}=80^{+26}_{-22}$ ${\rm km\ s}^{-1}$and $\sigma_{RV,cl}=48^{+15}_{-15}$ ${\rm km\ s}^{-1}$. 
    On the other hand, we found no third component in the Minni\,19 RV distribution because the simulated stars of the disk and bulge reproduced the observations well. The cluster velocity distribution might overlap the disk or the bulge, or there might be no cluster at all. For Minni\,10, we were unable to conclude if the distribution has a component associated with a cluster. Different tests could be applied to see if these results remain the same, and better-quality data may be needed, as Minni\,10 was the cluster with the largest RV error, where we had to remove a large number of low-quality stellar spectra from the analysis (see Figure \ref{fig:Minni10RV}). 
    
    For the case of Minni\,07, for which we confirm the detection of a RV peak, we investigated it further to find other evidence to confirm its GC nature. We analyzed the distribution of [Fe/H] derived from the MUSE spectra and discovered that Minni\,07 is significantly different from the bulge and disk at [Fe/H]$\approx-0.27$, which we assumed as the cluster metallicity. We also derived an extinction of $A_V=1.5$ mag, consistent with a previous determination. Using multiband CMDs and isochrones for Minni\,07 with a fixed metallicity and extinction derived here and a distance of 6.8\,kpc taken from red clump analysis by \citet{Minniti+2017c}, we compared the CMDs with isochrones of different ages (Fig. \ref{fig:minni07isoc}). In order to identify the general shape of the Minni 07 CMD, we color-coded the stars with their RV information. We took the normalized height of the RV Gaussian function fitted to Minni 07, where 1 means that the star is right at the RV peak, and lower numbers follow a normalized Gaussian function away from the peak down to zero. The youngest 2 Gyr old isochrone clearly has a much brighter main-sequence turn-off that does not fit any star. The 6 Gyr old isochrone is borderline, which is best shown in the three bottom panels. The older isochrones are the smallest difference in magnitude. We can therefore rule out an age as young as 2 Gyr and conclude that Minni 07 is probably older than 6 Gyr, but the current data are not precise enough to define whether it is 10 or 14 Gyr old. The visual inspection of the CMDs and isochrones led to the conclusion that this is likely an old and metal-rich GC, similar to Minni\,23 \citep{Dias+2022}.

     Although we found a component that might be associated with a GC for Minni\,23 and Minni\,07, we note that their velocity dispersion is too high for a typical globular cluster. However, these velocity dispersion values have a large error bar and can be consistent with a real GC within $2\sigma$ for the two Minni clusters. These large errors might arise from contamination of field stars whose RV distributions overlap with that of the potential low-mass cluster. \citet{Carvajal+2022} performed a similar analysis of a similar dataset, but for that particular cluster, the RV peak was far away from the RV bulge+disk distribution. They reported an error of the mean cluster RV of 0.8 $km s^{-1}$ based on 55 cluster member stars, which means a velocity dispersion of 5.9 $km s^{-1}$. This dispersion is about that of the typical uncertainties of individual RV measurements, as we discussed in Section 3. This proves that when the RV peak is detached from any foreground contamination, it is possible to recover the cluster. We studied intrinsically challenging cases and we were still able to detect a signature of a cluster within a particular RV range. However, we cannot assume that the resulting velocity dispersion is realistic for the reasons discussed above, combined with potential systematic errors from the models.
     Further high-resolution observations for Minni\,23 and Minni\,07 would allow us to obtain better RV measurements to detect the peak more precisely and would in addition enable us to study the chemical abundances of these clusters and the star membership selection.
    
	The search for more GCs, such as Minni\,07, using Gaia DR4 spectroscopic data that will be released in late 2026 in combination with MUSE data for the other candidate GCs located deep in the bulge might uncover some of the bulge GCs that are challenging to observe and reveal more missing clusters of the MW.

\section*{Data availability}
The table containing the coordinates, atmospheric parameters, and radial velocities for all stars in the four MUSE datacube is only available in electronic form at the CDS via anonymous ftp to cdsarc.u-strasbg.fr (130.79.128.5) or via http://cdsweb.u-strasbg.fr/cgi-bin/qcat?J/A+A/
\begin{acknowledgements}
      P.R.C. acknowledges support by ANID-FONDECYT iniciación grant No. 11221366. 
      B.D. acknowledges support by ANID-FONDECYT iniciación grant No. 11221366 and from the ANID Basal project FB210003.  
      D.M. acknowledges support by ANID Fondecyt Regular grant No. 1220724, and by the BASAL Center for Astrophysics and Associated Technologies (CATA) through ANID grants ACE210002 and FB 210003.
      Based on observations collected at the European Southern Observatory under ESO programme 0101.D-0363(A). 
       We gratefully acknowledge data from the ESO Public Survey program ID 179.B-2002 taken with the VISTA telescope, and products from the Cambridge Astronomical Survey Unit (CASU).
        This work has made use of data from the European Space Agency (ESA) mission Gaia (\url{https://www.cosmos.esa.int/gaia}), processed by the Gaia Data Processing and Analysis Consortium (DPAC, \url{https://www.cosmos.esa.int/web/gaia/dpac/consortium}). Funding for the DPAC has been provided by national institutions, in particular the institutions participating in the Gaia Multilateral Agreement.
       This research uses services or data provided by the Astro Data Lab at NSF’s National Optical-Infrared Astronomy Research Laboratory. NOIRLab is operated by the Association of Universities for Research in Astronomy (AURA), Inc. under a cooperative agreement with the National Science Foundation 
      
\end{acknowledgements}

   \bibliographystyle{bibtex/aa} 
   \bibliography{Bibliography} 

   \clearpage
   \onecolumn
   \begin{appendix}
    \section{Extra figures}
    Figures for distributions of radial velocities for Minni 10, Minni 19, and Minni 23 after applying the error cut, KS-test of the observations and the model of fields stars, and MCMC for the case of Minni 23.

\begin{figure}[h]
    \centering
    \includegraphics[width=0.30\textwidth]{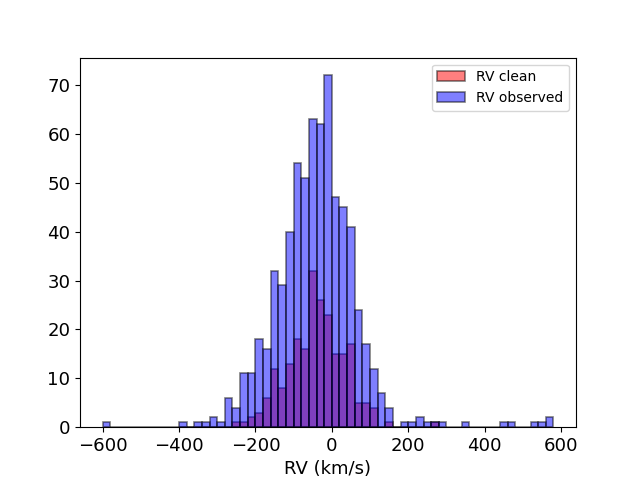}
    \includegraphics[width=0.30\textwidth]{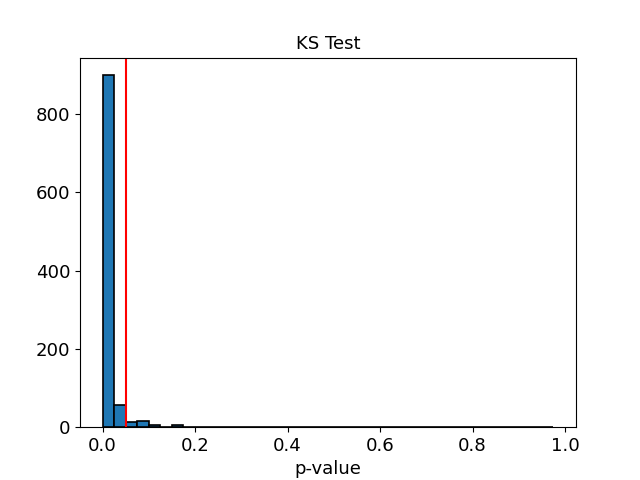}
    \includegraphics[width=0.30\textwidth]{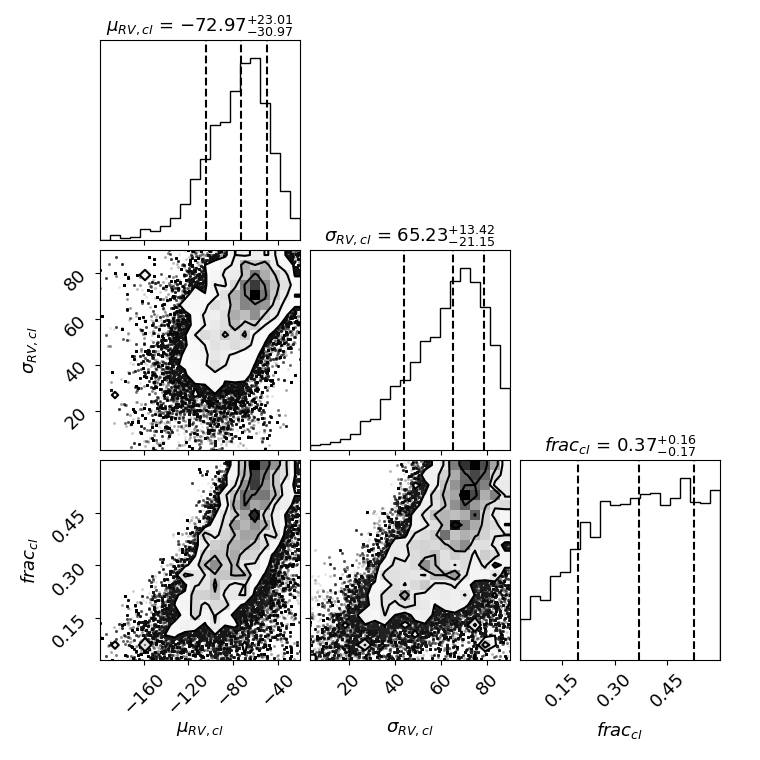} 
    \caption{Similar to Figs. \ref{fig:RVcleaned},
    \ref{fig:KStest}, and \ref{fig:mcmc Minni23} for Minni\,10.}
    \label{fig:Minni10RV}
\end{figure}

\begin{figure}[h]
    \centering
    \includegraphics[width=0.30\textwidth]{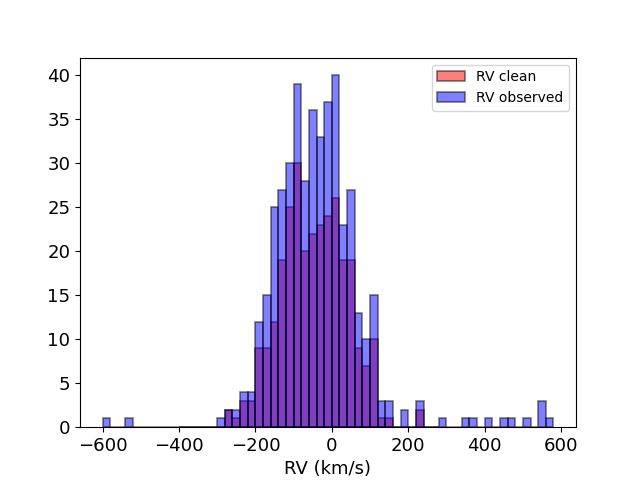}
    \includegraphics[width=0.30\textwidth]{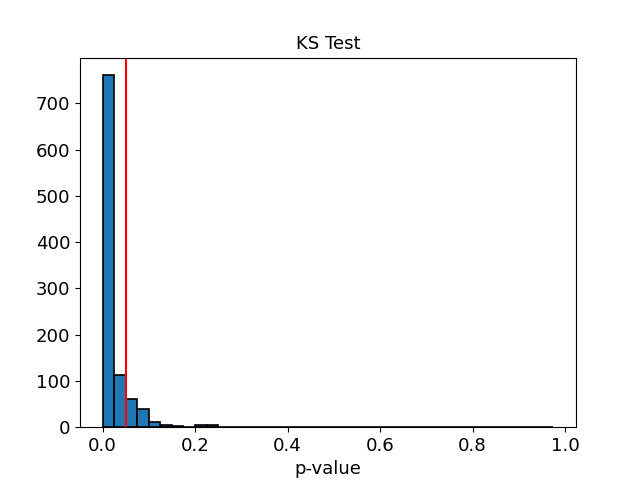}
    \includegraphics[width=0.30\textwidth]{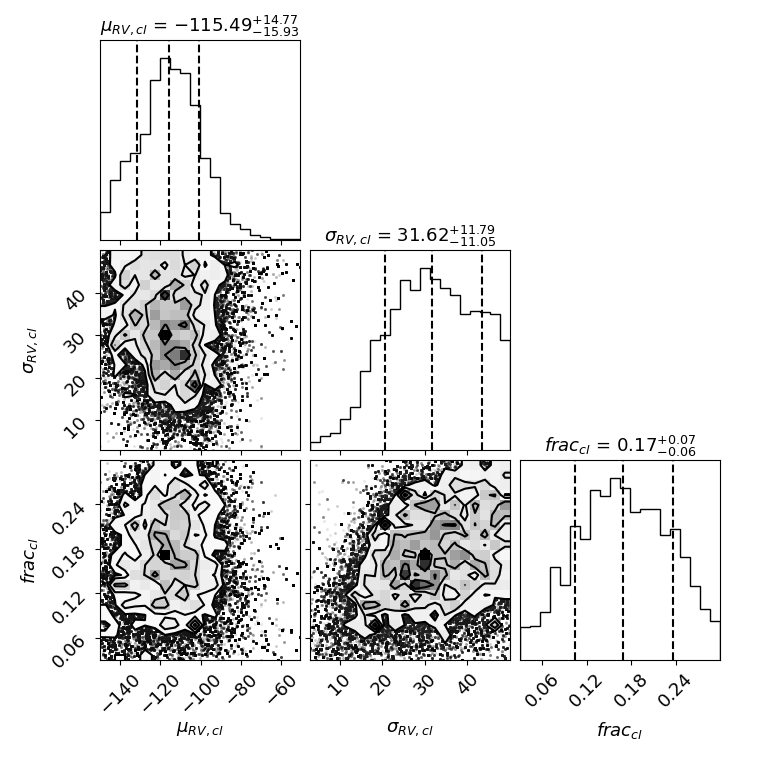}   
    \caption{Similar to Figs. \ref{fig:RVcleaned},
    \ref{fig:KStest}, and \ref{fig:mcmc Minni23} for Minni\,23.}
\end{figure}

\begin{figure}[h]
    \centering
    \includegraphics[width=0.40\textwidth]{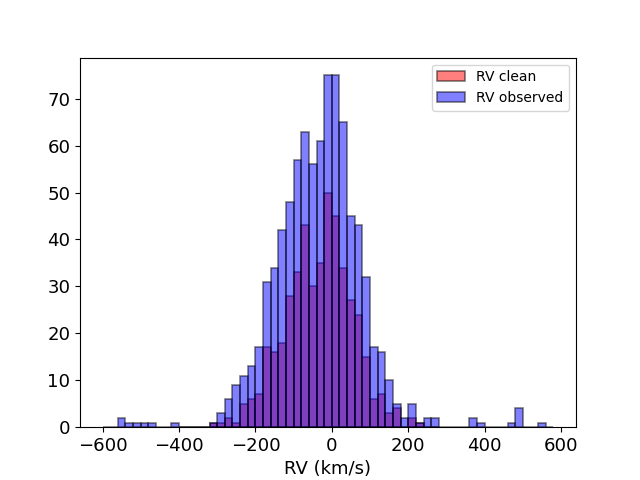}
    \includegraphics[width=0.40\textwidth]{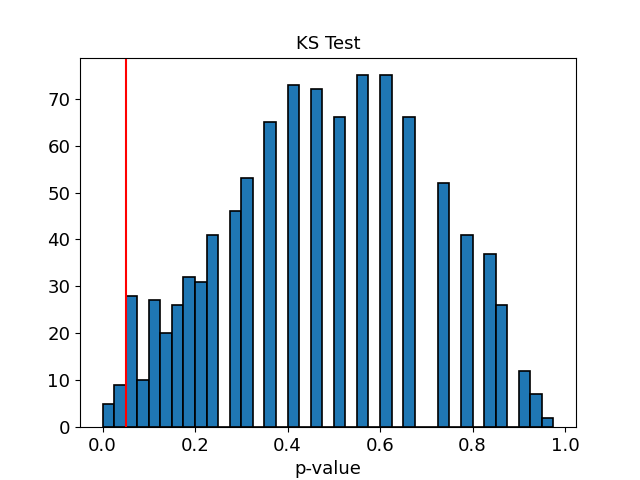}
    \caption{Similar to Figs. \ref{fig:RVcleaned} and
    \ref{fig:KStest} for Minni\,19.}
\end{figure}

\end{appendix}

\end{document}